\documentclass{revtex4}

\usepackage{epsf,graphicx,subfigure,amsmath,amssymb,amsfonts}

\begin{document}

\newcommand{\avg}[1]{\left\langle{#1}\right\rangle}
\newcommand{\davg}[1]{\left\langle\!\left\langle{#1}\right\rangle\!\right\rangle}
\newcommand{\ovl}[1]{\overline{#1}}
\renewcommand{\l}{\left}
\renewcommand{\r}{\right}
   \newcommand{\nvec}[1]{\stackrel{\rightarrow}{#1}}
   \newcommand{\vecbm}[1]{\mbox{\boldmath#1}}
   \newcommand{\vecb}[1]{\mbox{\bf#1}}
   \newcommand{\triplint}{\int\rule{-3.5mm}{0mm}\int\rule{-3.5mm}{0mm}\int}
   \newcommand{\doublint}{\int\rule{-3.5mm}{0mm}\int}
   \newcommand{\lora} {\boldmath$\longrightarrow$}
\author{D.H.E. Gross}

\affiliation{Hahn-Meitner-Institut
  Berlin, Bereich Theoretische Physik,Glienickerstr.100\\ 14109
  Berlin, Germany and Freie Universit{\"a}t Berlin, Fachbereich
  Physik; \today}

\title{ON THE MICROSCOPIC FOUNDATION OF THERMO-STATISTICS}

\pacs{numbers here}

\begin{abstract}
  The most complicated phenomena of equilibrium statistics, phase
  separations and transitions of various order and critical phenomena,
  can clearly and sharply be seen even for small systems in the
  topology of the curvature of the microcanonical entropy
  $S_{B}(E,N)=\ln[W(E,N)]$ (Boltzmann's principle $BP$) as function of
  the conserved energy, particle number etc.. Consequently, $BP$
  allows to establish the link toward their microscopic origin and the
  study of the way how interacting many-body systems organize into
  phase-transitions. Also the equilibrium of the largest possible
  interacting many-body systems like self-gravitating systems is
  described to great extend by the topology of the entropy surface
  $S_{B}(E,N,\vecbm{$L$})$ where \vecbm{$L$} is the angular momentum.
  Conventional (canonical) statistical mechanics describes only a
  small section out of all equilibrium phenomena in nature and only in
  cases where the so-called ``thermodynamic limit'' applies
  (homogeneous phases of ``infinite'' systems interacting with
  short-range interactions). In this paper I present two examples of
  phase transitions of first order which are of fundamental
  importance: the liquid to gas transition in a small atomic cluster
  and the condensation of a rotating self-gravitating system into
  single stars or into multi-star systems like double stars an rings.
  Such systems cannot be addressed by ordinary canonical
  thermo-statistics.  I also give a geometric illustration how an
  initially non-equilibrized ensemble approaches the microcanonical
  equilibrium distribution.
\end{abstract}

\maketitle
\section{Introduction}
Many theorems we are used to in conventional macroscopic (canonical)
thermo-statistics are wrong when statistical mechanics addresses small
or other non-extensive systems. Here a revision of the fundamentals is
demanded.

Phase separation of normal systems and also in general the equilibrium
of closed non-extensive systems are not described by the canonical and
grand-canonical ensembles. Only the microcanonical ensemble based on
Boltzmann's entropy $S_{B}=\ln{W}$, with $W(E)=\epsilon_0
tr[\delta(E-H)]$, the number of classical or quantum states, describes
correctly the unbiased uniform filling of the energy-shell in
phase-space.

Various ensembles like the (grand-)canonical are equivalent to the
microcanonical ensemble only if the system is infinite and
homogeneous, i.e. in a pure and homogeneous phase. Only then exists a
one to one mapping from the conserved mechanical observables as energy
$E$, particle number $N$, eventually angular-momentum \vecbm{$L$} and
others to the intensive variables temperature $T$, chemical potential
$\mu$ and eventually rotational frequency $\omega$ and so on.
Otherwise the (grand-)canonical ensembles do not reflect the
equilibrized phase-space distribution of a closed ergodic Hamiltonian
system~\cite{gross124,gross140,gross158,gross174} see also Barr\'{e}
et al~\cite{barre01}. (Grand-)canonical potentials are non-analytic at
phase transitions whereas $S_B(E,N,\cdots)$ remains multiply
differentiable also there.

This program is far from only academic interest and is deeply demanded
in many fields of condensed matter. A pseudo Riemannian geometry must
replace Ruppeiner's Riemannian geometry of
fluctuations~\cite{ruppeiner95,andresen96}. It leads to negative heat
capacities at phase-separation which cannot be explained in any
canonical formalism and which are well documented experimentally
c.f.~\cite{thirring70,chbihi95,lyndenbell95a,lyndenbell99,gross171,gross172,schmidt00,dAgostino00,schmidt01}
as also postulated theoretically since long
cf.~\cite{gross95,gross158,gross150,gulminelli99a,casetti99a,ispolatov01,gross181,gross187,gross190,gross191}.

Besides conventional extensive systems in the thermodynamic limit,
this formulation of thermo-statistics addresses additionally the
following objects:
\begin{itemize}
\item Astro-physical systems with their typical negative heat capacity
  are clearly outside of any canonical
  approach~\cite{thirring70,lyndenbell99,casetti99a,ispolatov01,gross187,gross190,gross191}.
\item The same is true for the original goal of Thermodynamics, the
  description of phase
  separation~\cite{gross174,gross189,gross176,gross182,gross177,gross138},
\item and of course small systems like excited nuclei, atomic clusters
  etc., which are addressed more recently, where many experimental
  results are now
  cumulating~\cite{chbihi95,lyndenbell95a,gross171,gross172,schmidt00,dAgostino00,schmidt01},
  and which are exotic from the canonical point of view
  c.f.section(\ref{surfS}).
\end{itemize}
\section{Boltzmann's entropy \vecbm{$S_B$} vs. equipartition entropy 
\vecbm{$S_{equip.}$}}
The microcanonical ensemble is defined by the uniform (completely
unbiased) probability distribution on the energy manifold
{${\cal{E}}$}:
\begin{eqnarray}
\hat{P}&=&\frac{\delta(E-\hat{H})}{tr[\delta(E-\hat{H})]}
\label{microprob}\\
tr[\hat{A}]&:=&\int{d^{6N}q\;A(\nvec{q})}\\
\nvec{q}&\equiv&\nvec{x},\nvec{p}
\end{eqnarray}
Boltzmann's entropy is then (Boltzmann' principle):
\begin{equation}
S_B=\ln{\left\{\frac{\epsilon_0}{N!(2\pi\hbar)^{3N}}\int{d^{6N}q\;
\delta[E-H(\nvec{q})]}\right\}}\label{boltzmprinciple}
\end{equation}
That is the definition of entropy we used in all our previous work.

There may be another entropy: Here we follow the book of
Berdichevsky~\cite{berdichevsky97} for a little while:
\begin{eqnarray}
<q_i\frac{\partial H}{\partial q_i}>&=&\frac{\int{d^{6N}q\;
\delta[E-H(\nvec{q})]q_i}\frac{\partial H}{\partial q_i}}
{\int{d^{6N}q\;\delta[E-H(\nvec{q})]}}\label{equi1}
\\
&\equiv&\frac{\int{d^{6N}q\;\Theta[E-H(\nvec{q})]}}
{\int{d^{6N}q\;\delta[E-H(\nvec{q})]}}\label{equi2}\\
&\equiv&<q_k\frac{\partial H}{\partial q_k}>\\
&=:&T_{equipart.}\label{equipartLaw}
\end{eqnarray}
where the integral over the $\delta$-function in eq.(\ref{equi1}) was
transformed into an integral over the $(6N-1)$-dim. energy surface and
then for the step from eq.(\ref{equi1}) to (\ref{equi2}) the law of
Gauss-Ostogradski was used. As eq.(\ref{equi2}) is independent of
which d.o.f is taken, this is the equipartition theorem for a finite
system.

One is now tempted to define an equipartition entropy $S_{equipart.}$
\begin{eqnarray}
\frac{dS_{equipart.}}{dE}&:=&\frac{1}{T_{equipart.}}\\
&=&\frac{\int{d^{6N}q\;\delta[E-H(\nvec{q})]}}
{\int{d^{6N}q\;\Theta[E-H(\nvec{q})]}}\\
S_{equipart.}&=
&\ln{\left\{\frac{1}{N!(2\pi\hbar)^{3N}}\int{d^{6N}q\;
\Theta[E-H(\nvec{q})]}\right\}}.
\label{S_2}
\end{eqnarray}
However, the microcanonical probability $\hat{P}$ is still (c.f.
eqs.(\ref{microprob},\ref{equi1}))
\begin{eqnarray}
\hat{P}&=&\frac{\delta(E-\hat{H})}{tr\delta(E-\hat{H})}\\
\lefteqn{\mbox{Thus :}}\nonumber\\
S_{equipart.}&\ne&-tr[\hat{P}\ln(\hat{P})],
\end{eqnarray}
because $\hat{P}= 0$ for $E\ne H$ conform to
eq.(\ref{boltzmprinciple}) and $S_B$. $S_{equipart.}$ has no
interpretation within information theory: It depends on regions in
phase space inaccessible to the system when the conservation of energy
holds.  Therefore, we decided at variance to
ref.~\cite{berdichevsky97} to use the original Boltzmann entropy
eq.(\ref{boltzmprinciple}). The relative difference between both is
$\propto 1/N$.  For medium large systems with short range interactions
this difference is negligible also in comparison to the surface
entropy $\propto N^{-1/3}$ which is the main origin of the positive
curvature discussed in section \ref{surfS}. Thus here the
equipartition law (\ref{equipartLaw}) holds but the $T_{equipart.}$
differs from $T=(\partial S_B/\partial E)^{-1}$ by a term of the order
$1/N$.
\section{Phase-transitions microcanonically and grand-canonically}
\begin{figure}[h]
\begin{center}
\includegraphics*[bb =  53 14 430 622, angle=-90, width=9cm,  
clip=true]{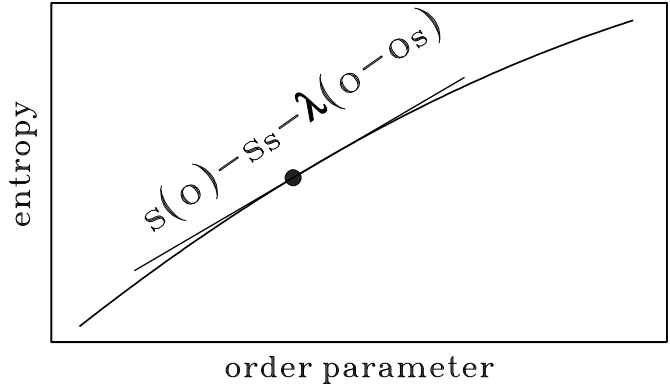}
\caption{Concave entropy with single stationary point $o_s$. In the limit
  $N\to\infty$ this gives a one-to-one mapping of $E_s,N_s$ to
  $T,\mu$. The grand-canonical $Z(T,\mu)$ has the same analytic
  properties as the microcanonical $W(E,N)$.  This is the condition
  for a pure phase in the conventional Yang-Lee theory~\cite{yang52}.
  Consequently, we define a pure phase microcanonically also for a
  finite system by a region of negative maximum curvature of
  $S(E,N,\cdots)$.\label{concave}}
\end{center}
\end{figure}

For an orientation, how to generalize the concept of phase transitions
to microcanonical {\em finite} systems we start with the Yang-Lee
theory for the grand-canonical ensemble of large, homogeneous systems.
The grand-canonical ensemble for $V\to\infty$:\\
\begin{eqnarray}
 Z(T,\mu,V)&=&\doublint_0^{\infty}{\frac{dE}{\epsilon_0}\;dN\;e^{-[E-\mu
N-TS(E)]/T}}\label{grandcanon}\\
&=&\frac{V^2}{\epsilon_0}\doublint_0^{\infty}{de\;dn\;e^{-V[e-\mu
n-T{\mbox{\boldmath$\scriptstyle s(e,n)$}}]/T}}\label{grandsum}.\\
&\approx&\hspace{1cm}e^{\mbox{ cons.+lin.+quadr.}}\nonumber
\end{eqnarray}\\
We put the linear term to 0 : If $s(e,n)$ is concave, c.f.
figure(\ref{concave}), then there is a single stationary point
$e_s$,$n_s$ with
\begin{eqnarray}
\frac{1}{T}&=&\left.\frac{\partial S}{\partial E}\right|_{e_s,n_s}\nonumber\\
\frac{\mu}{T}&=&-\left.\frac{\partial S}{\partial N}\right|_{e_s,n_s}
\nonumber
\end{eqnarray}
and the Laplace transform can be done in Gaussian approximation to get
the free energy density:
\begin{eqnarray}
\frac{-T\ln[Z(T,\mu,V)]}{V}&\to& e_s-\mu n_s-Ts_s\\
\lefteqn{\hspace{-3cm}+\frac{T\ln{(\sqrt{(-\lambda_1)}
\sqrt{(-\lambda_2)})}}{V}+o(\frac{\ln{V}}{V})}
\end{eqnarray}

Consequently, in the fundamental microcanonical ensemble pure phases
are indicated by a negative maximum curvature, eigenvalue $\lambda_1$
of the Hessian matrix:
\begin{eqnarray}
{\cal{H}}(e,n)&=& \left(\begin{array}{cc}
\frac{\partial^2 s}{\partial e^2}& \frac{\partial^2 s}{\partial n\partial e}\\
\frac{\partial^2 s}{\partial e\partial n}& \frac{\partial^2 s}{\partial n^2}
\end{array}\right)\label{hessian}\\
\mbox{eigen-values: }\lambda_1\ge\lambda_2&\hspace{1cm}
\mbox{\lora eigenvectors :}\hspace{1cm}
{\boldmath\vecbm{$v$}_1,\vecbm{$v$}_2}\nonumber\\
{\boldmath\vecbm{$v$}_1}&=&\mbox{direction of order parameter }o
\end{eqnarray}
At phase transition, the grand-canonical partition function $Z(T,\mu)$
has a zero in the thermodynamic limit (Yang-Lee zero \cite{yang52}).
Here the grand-canonical energy distribution (the kernel of integral
(\ref{grandcanon})) is bimodal c.f. figure (\ref{na1000}). The pure
``liquid'' phase, at $e_1$ and the pure ``gas'' phase at $e_3$ coexist
at this temperature with equal probability. This is of course only
possible if $s(e,n)$ has a positive curvature between $e_1$ and $e_3$.
I.e. the general microcanonical condition for a phase transition of
first order is the occurrence of a positive maximal curvature
$\lambda_1$. This can now be generalized to finite systems.

\section{Equilibrium of small excited metal clusters \label{surfS}}
\begin{figure}[h]
\begin{center}
\includegraphics*[bb = 84 43 409 303, angle=-0, width=6.5cm,  
clip=tru]{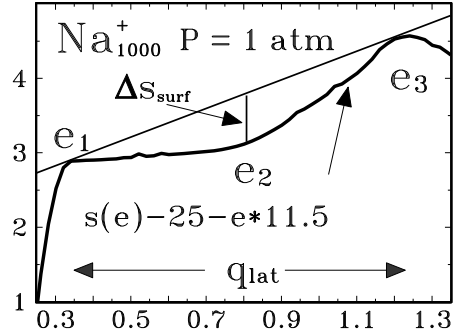}
\caption{Entropy per atom $s(e)$ as function of the energy $e$ per atom
  for a self bound cluster of $1000$ sodium atoms. \label{na1000} The
  double tangent between $e_1$ and $e_3$ is the concave hull of
  $s(e)$. Its derivative gives the Maxwell construction in the caloric
  curve $T(e)$.}
\end{center}
\end{figure}

In this section I will shortly discuss the statistical equilibrium of
a typical small self-bound system, a cluster of a few hundred
$Na$-atoms interacting by a realistic many-body (metallic) force at
{\em constant pressure of 1 atm.} i.e. with the constraint
$\frac{\partial S/\partial V}{\partial S/\partial E}=1 atm$. As
function of the excitation energy a typical convex intruder shows the
signals of a liquid to gas phase-transition of first order, c.f.
fig.(\ref{na1000}). The derivative of the concave hull of $S(E)$ gives
the Maxwell construction of the caloric curve $T(E)$. The depth
$\Delta s_{surf}$ of the intruder (measured from the concave hull) is
a measure of the surface tension c.f.\cite{gross174}. Table
(\ref{tabNa}) compares the transition temperature $T_{tr}$ (inverse
slope of the hull) in Kelvin, the latent heat per atom
($q_{lat}=e_3-e_1$) in e.V., the boiling entropy (the gain of entropy
when one atom is evaporated at $T_{tr}$), the surface entropy (the
depth of the intruder $\Delta s_{surf}$), the average number
$N_{surf}$ of surface atoms summed over all clusters, and finally the
surface tension $\sigma_{surf}/T_{tr}=s_{surf}N_0/N_{surf}$ per
surface atom divided by the transition temperature in dimensionless
units, with the known bulk values.

This figure shows clearly the necessary condition for a
phase-separation, the {\em bimodality}, the appearance of two equal
minima of the canonical free energy $f=e-Ts$ at $e_1$ and $e_3$, which
can only appear when the entropy $s(e)$ has a positive curvature and
correspondingly a negative heat capacity
\begin{equation}
c=-\frac{(\frac{\partial S}{\partial E})^2}{\frac{\partial^2S}{\partial E^2}}
\end{equation}
in between.
\begin{table}[h]
\begin{center}
\begin{tabular} {|c|c|c|c|c|c|} \hline 
&$N_0$&$200$&$1000$&$3000$&bulk\\ 
\hline 
\hline  
&$T_{tr} \;[K]$&$940$&$990$&$1095$&1156\\ \cline{2-6} 
&$q_{lat} \;[eV]$&$0.82$&$0.91$&$0.94$&0.923\\ \cline{2-6} 
{\bf Na}&$s_{boil}$&$10.1$&$10.7$&$9.9$&9.267\\ \cline{2-6} 
&$\Delta s_{surf}$&$0.55$&$0.56$&$0.45$&\\ \cline{2-6} 
&$N_{surf}$&$39.94$&$98.53$&$186.6$&$\infty$\\
\cline{2-6} 
&$\sigma/T_{tr}$&$2.75$&$5.68$&$7.07$&7.41\\ 
\hline
\end{tabular}
\end{center}
\caption{Comparison of the parameters of the liquid-gas transition of a 
cluster of $200$, $1000$ and $3000$ $Na$-atoms with those of the bulk 
at an external pressure of $1$atm.\label{tabNa}. Details 
c.f.~\cite{gross174}}
\end{table}
\section{Equilibrium of self-gravitating and rotating systems}
Self-gravitation leads to a non-extensive potential energy $\propto
N^2$.  No thermodynamic limit exists for $S/N$ and no canonical
treatment makes sense. At negative total energies these systems have a
negative heat capacity.  This was for a long time considered as an
absurd situation within canonical statistical mechanics with its
thermodynamic ``limit''. However, within our geometric theory this is
a simple example of the pseudo-Riemannian topology of the
microcanonical entropy $S(E,N)$ provided non-gravitational physics at
high densities is excluded.  The fundamental importance of
self-gravitating systems calls urgently for an abolishment of several
un-physical axioms used in conventional
statistics~\cite{gross189,gallavotti99}.

\begin{figure}[h]
\begin{center}
  \includegraphics[width=6.5cm,angle=-90]{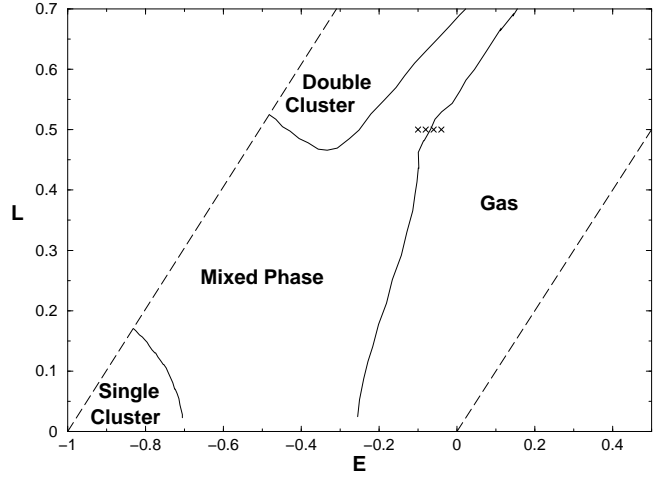}
\caption{\label{phased}Phase diagram of a microcanonical, rotating 
  self-gravitating system [topology of the Hessian,
  eq.(\ref{hessian})], in the $(E,L)$-plane. The dashed lines $E-L=1$
  (left) and $E=L$ (right) delimit the region where the Hessian was
  calculated. In the mixed phase the largest curvature is
  $\lambda_1>0$}
\end{center}
\end{figure}

\begin{figure}[h]
\begin{center}
\subfigure[$E=-0.72$,
$L=0.4$]{\scalebox{.53}{\includegraphics{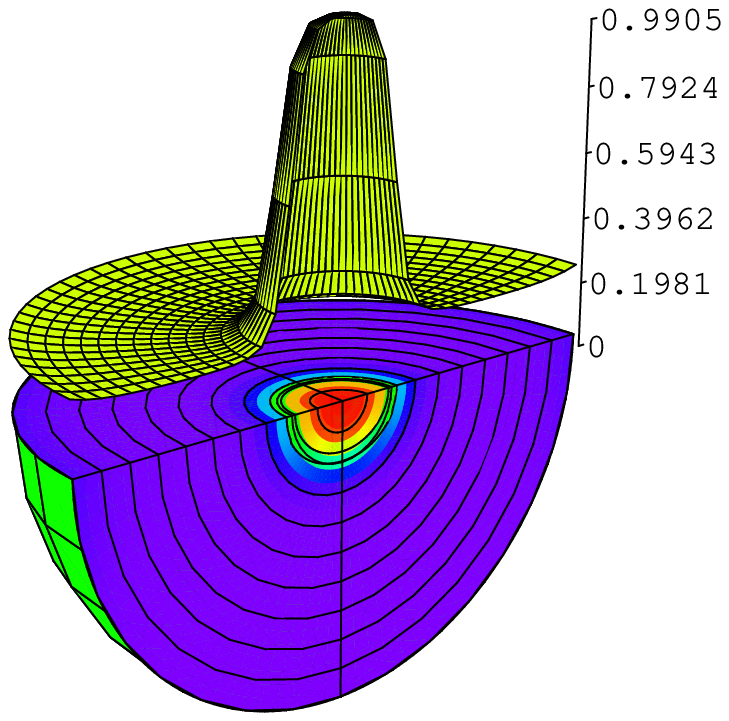}}}
\subfigure[$E=-0.9$,
$L=0.5$]{\scalebox{.53}{\includegraphics{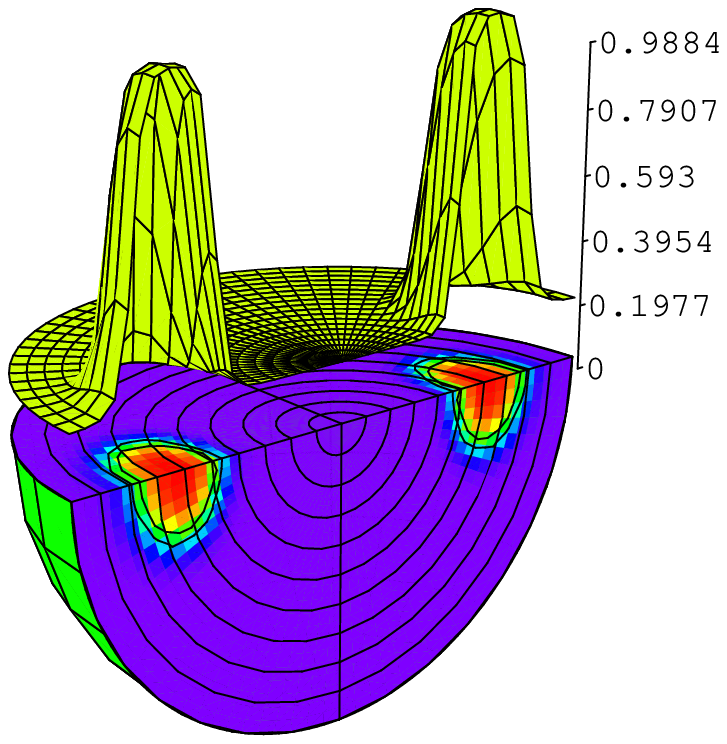}}}
\subfigure[$E=-0.06$,
$L=0.4$]{\scalebox{.53}{\includegraphics{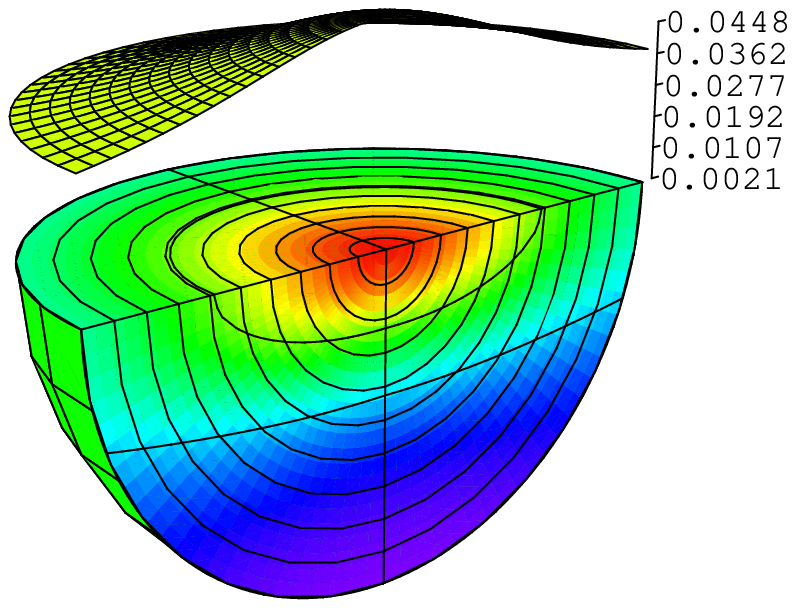}}}
\subfigure[$E=-0.42$,
$L=0.5$]{\scalebox{.53}{\includegraphics{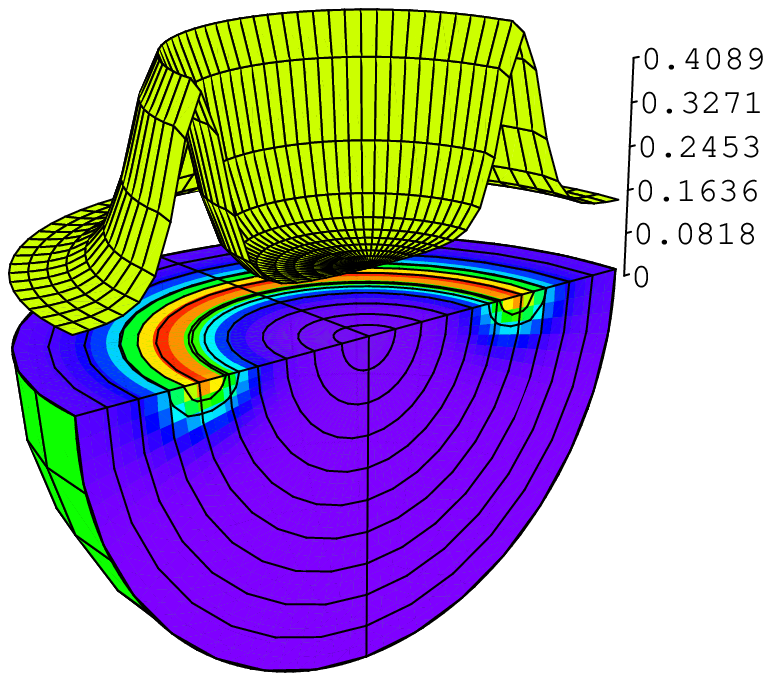}}}
\caption{\label{distri} Color online: Examples of stationary distributions
  $c(\boldsymbol{x})$ occurring inside our spherical box. The angular
  momentum is along the vertical axis. Shown are the contour plot and,
  above it, the density profile: (a) ``single star'', (b) ``double
  star'', (c) ``disk'', (d) ``ring''.  C.f.: Physical Review
  Letters--July 15, 2002, cover-page, calculations by Votyakov et.
  al.~\cite{gross187}}
\end{center}
\end{figure}

To prove the power of our geometric theory of statistical mechanics I
show in figure (\ref{phased}) the global phase diagram of a system of
$N$ self-gravitating particles as function of total energy $E$ and
given total angular-momentum \vecbm{$L$}.  As can clearly be seen, our
geometric formulation of statistical mechanics gives a quite realistic
picture of the phase-transition into mono-star or at higher
\vecbm{$L$} into multi star configurations even though this is far
beyond any thermodynamic ``limit''. In the region of mixed phases one
has at any $E$,\vecbm{$L$} point a separation into mono-star,
double-star, ring-like, and even more complicated inhomogeneous
configurations and the all space-filling gas phase. In this mixed
phase the entropy $S(E,N)$ has at least one positive eigen-curvature
and the system has a negative specific heat (more generally, a
negative susceptibility), the characteristic signal of a phase
transition of first order.  More details are in
refs~\cite{gross187,gross190,gross191}.
Figure (\ref{distri}) gives a few examples of the rich and realistic
configurations of the {\em statistical equilibrium} of such a
self-gravitating and rotating many-body system. This demonstrates the
superiority of our geometric statistics over any conventional
canonical theory.  These systems are far larger than any conventional
thermodynamic ``limit'' can dream of.
\clearpage
\section{Geometric illustration of the approach to equilibrium and 
  the Second Law} 

As Thermodynamics describes a $N$-body systems by a few control
parameters $M$ only, which are much less than the total number $6N$ of
degrees of freedom, it gives only {\em probabilistic} answers (the
{\em average} of some observable $<\!O\!>$ over the whole ensemble of
all systems with the same values of the $M$ control-parameters).
Therefore, Thermodynamics describes the evolution of the whole
ensemble. There is also a geometric interpretation of the evolution of
a non-equilibrized ensemble to the microcanonical uniform filling of
the energy-shell in the $6N$-dim.  phase space.

Even though every trajectory spreads over the available phase space
and returns after $t_{Poincarre}$, different points of the manifold
have different $t_{Poincarre}$ which are normally incommensurable.
I.e. the ensemble spreads irreversibly over the accessible
phase-space.

\begin{figure}[h]
\begin{minipage}{6cm}
\begin{center}$V_a$\hspace{2cm}$V_b$\end{center}
\includegraphics*[bb = 0 0 404 404, angle=-0, width=5.7cm,
clip=true]{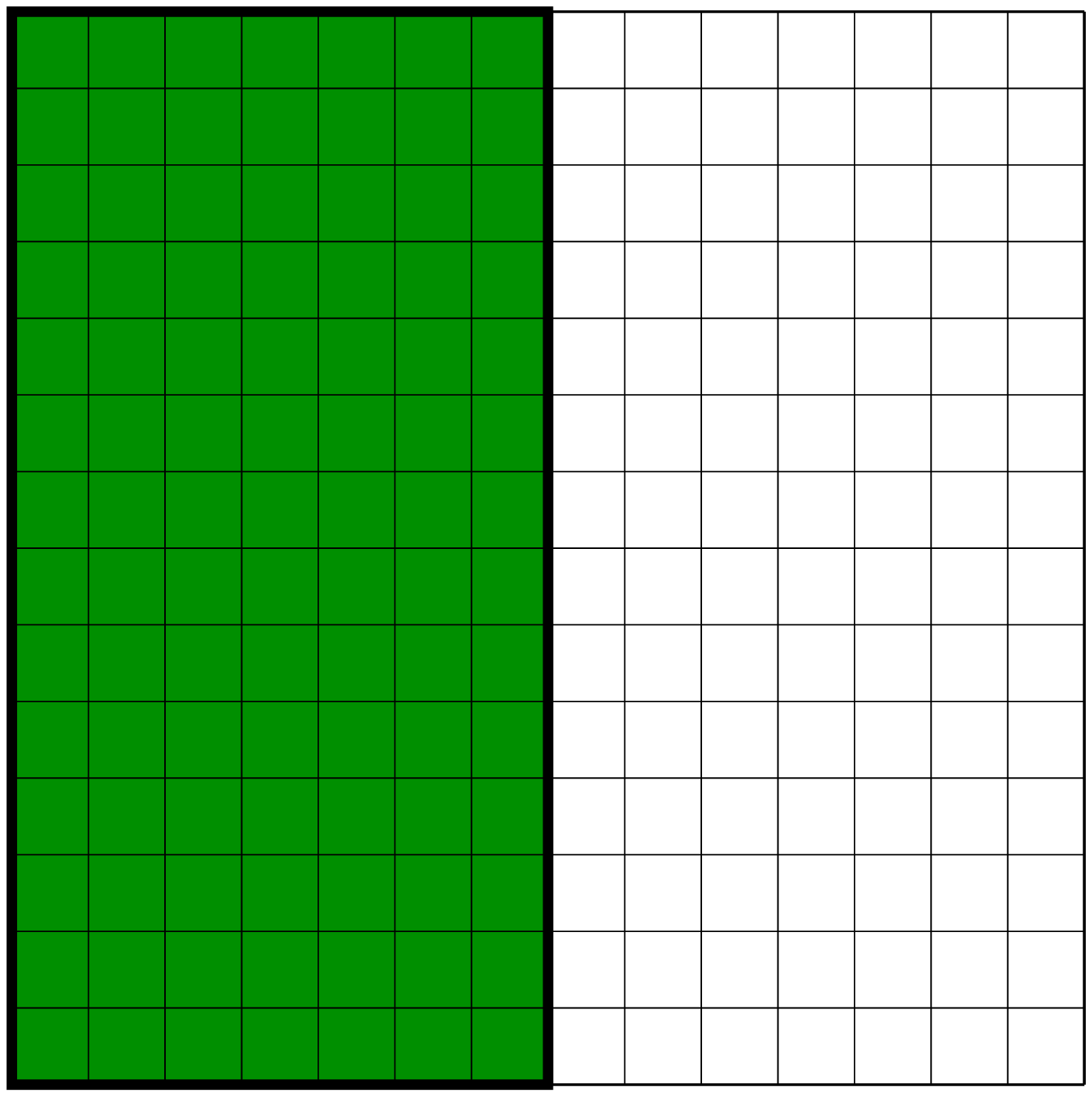}\begin{center}${\cal{M}}(t<t_0)$\end{center}
\end{minipage}$\longrightarrow$\begin{minipage}[h]{6cm}
\begin{center}$V_a+V_b$\end{center}\vspace{-0.8cm}
\includegraphics*[bb = 0 0 490 481, angle=-0, width=6.9cm,
clip=true]{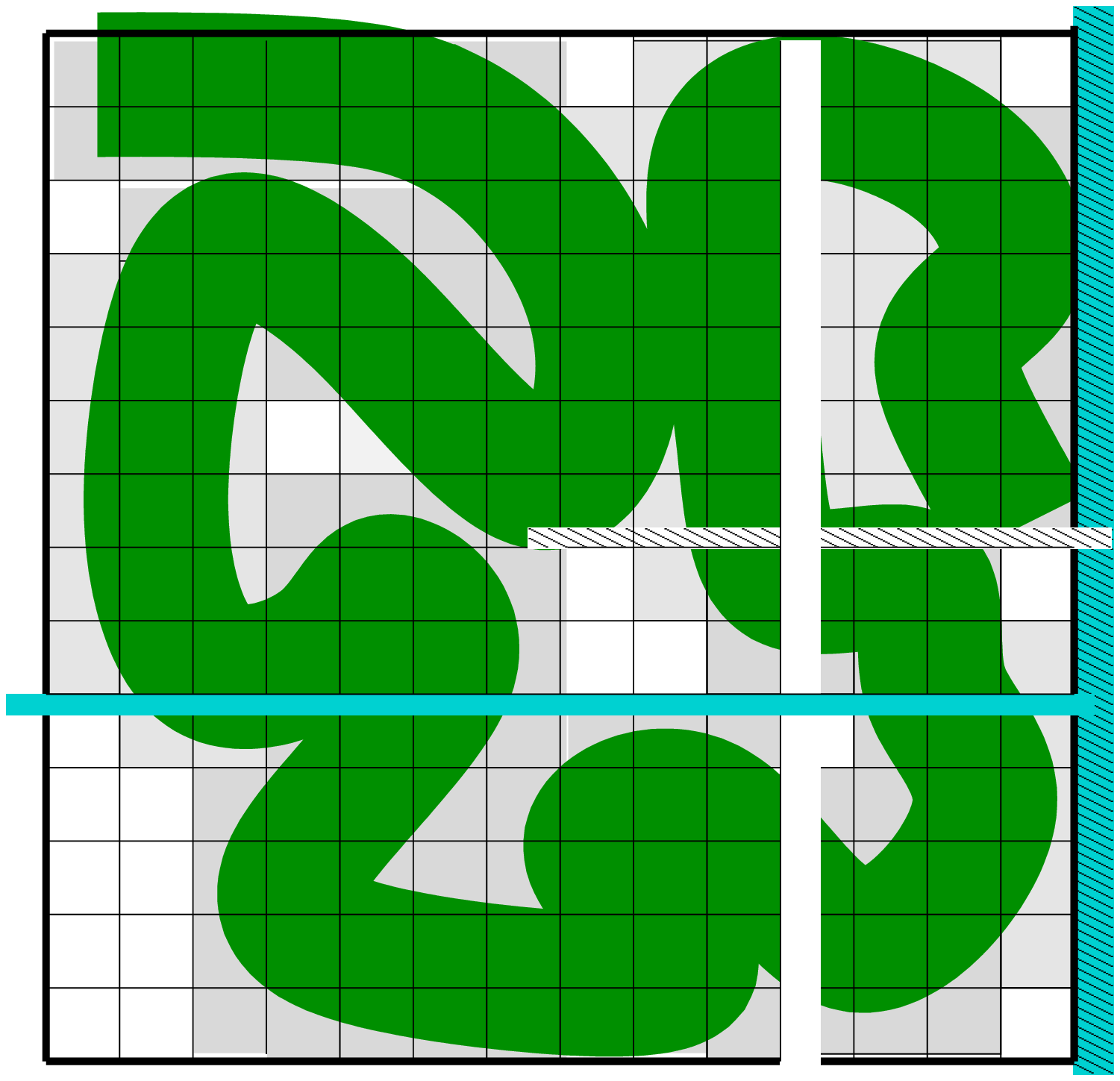}\begin{center}${\cal{M}}(t>t_0)$\end{center}
\end{minipage}
\caption{At time $t\le t_0$ the system is assumed to be equilibrized and
  be represented by the compact microcanonical set in phase-space
  ${\cal{M}}(t_0)={\cal{E}}(V_a)$, left side. At times $t_0$ the
  accessible volume is suddenly enlarged to $V_a+V_b$. The system
  develops into an increasingly folded but non-crossing
  ``spaghetti''-like distribution ${\cal{M}}(t)$ in phase-space with
  rising time $t>t_0$ after opening the volume $V_b$. The right figure
  shows only the early form of the distribution. At much later times
  it will become more and more fractal (Gibbs ink-lines) and finally
  dense in the new larger phase space.  The grid illustrates the boxes
  or size $\delta^{6N}$ of the box-counting method.  All boxes which
  overlap with ${\cal{M}}(t)$ contribute to the box-counting volume
  and are shaded gray.  Their number is $N_\delta$. Then the
  box-counting area is $=N_\delta \delta^d$ where $d$ is the dimension
  of the microcanonical manifold. \label{spaghetti}}
\end{figure}

Due to the redundancy of the information given by the few ($M$)
control-parameters one cannot distinguish the distribution in
phase-space from its direct neighborhood. Therefore, in the case of a
strongly folded (eventually fractal) non-equilibrium phase-space
distribution Boltzmann's entropy is the area of the closure of the
distribution~\cite{gross183}. The area of the closure can be
calculated by box-counting~\cite{falconer90,gross183} c.f.
fig.(\ref{spaghetti}). Mathematically, the area of the closure is
obtained in the limit of box-sides $\delta\to 0$. However, by several
reasons this limit should not be taken in physical applications (see
below).  This figure illustrates also how a non-equilibrium ensemble
develops in time, and how for a mixing dynamics it becomes more and
more dense in the larger available phase-space, so that the area of
its closure $e^{S(t)}$ approaches the larger area of the new
microcanonical ensemble $\propto(V_a+V_b)^N$.  This is the {\em
  geometric meaning of the Second Law of Thermodynamics}.  Due to the
inherent redundant information given by Thermodynamics, represented
here by a finite, non-zero, resolution $\delta$ of the box-counting,
this is achieved in a finite equilibration time~\cite{gross192}.
\section{Conclusion}

The {\em geometric interpretation of classical equilibrium Statistical
  Mechanics}~\cite{gross186} by Boltzmann's principle
(\ref{boltzmprinciple}) offers an extension also to the equilibrium of
non-extensive systems. In more fundamental, axiomatic terms, it opens
the application of Thermo-Statistics to ``non-simple'' systems which
are not (homogeneus) fluids or in contact with ideal gases. Surprisingly, 
but also understandably, this is still an open problem c.f.
ref.~\cite{uffink01} page 50 and page 72.

Because microcanonical Thermodynamics as a macroscopic theory controls
the system by a few, usually conserved, macroscopic parameters like
energy, particle number, etc. without fixing all $6N$ degrees of
freedom, it is an intrinsically probabilistic theory. It describes all
systems with the same control-parameters simultaneously. If we take
this seriously and avoid the so called thermodynamic limit
($\lim_{V\to\infty, N/V=\rho}$), the theory can be applied to small
systems but even to the really large, usually {\em inhomogeneous},
self-gravitating systems, c.f.\cite{gross187,gross190,gross191}.

Within the new, extended, formalism several principles of traditional
Statistical Mechanics turn out to be violated and obsolete. E.g. at
phase-separation (at negative heat capacity) heat (energy) can flow
from cold to hot~\cite{gross189}.  Or phase-transitions can be
classified unambiguously in astonishingly small systems. These are by
no way exotic and wrong conclusions. On the contrary, many experiments
have shown their validity.  I believe this approach gives a much
deeper insight into the way how many-body systems organize themselves
than any canonical statistics is able to.  The thermodynamic limit
clouds the most interesting region of Thermodynamics, the region of
inhomogeneous phase-separation.

Because of the only {\em one} underlying axiom, Boltzmann's principle
eq.(\ref{boltzmprinciple}), the {\em geometric
  interpretation}~\cite{gross186} keeps statistics most close to
Mechanics and, therefore, is more transparent. The Second Law ($\Delta
S\ge 0$) can even be shown to be valid in {\em closed, small} systems
under quite general dynamical conditions~\cite{gross192}.


\begin{thebibliography}{10}

\bibitem{gross124}
D.H.E. Gross and R.~Heck.
\newblock What is wrong with the Bethe formula ? - measurable differences
  between the grandcanonical and microcanonical ensemble.
\newblock {\em Phys. Lett.B}, 318:405--409, 1993.

\bibitem{gross140}
P.A. Hervieux and D.H.E. Gross.
\newblock Evaporation of hot mesoscopic metal cluster.
\newblock {\em Z. Phys.D}, 33:295--299, 1995.

\bibitem{gross158}
D.H.E. Gross and M.E. Madjet.
\newblock Microcanonical vs. canonical thermodynamics.
\newblock {\em http://xxx.lanl.gov/abs/cond-mat/9611192}, 1996.

\bibitem{gross174}
D.H.E. Gross.
\newblock {\em Microcanonical thermodynamics: Phase transitions in ``Small''
  systems}, volume~66 of {\em Lecture Notes in Physics}.
\newblock World Scientific, Singapore, 2001.

\bibitem{barre01}
Julien Barr\'e, David Mukamel, and Stefano Ruffo.
\newblock Inequivalence of ensembles in a system with long range interactions.
\newblock {\em Phys. Rev. Lett.},
  87:030601//http://arXiv.org/abs/cond--mat/0102036, 2001.

\bibitem{ruppeiner95}
G.~Ruppeiner.
\newblock Riemannian geometry in thermodynamic fluctuation theory.
\newblock {\em Rev.Mod.Phys.}, 67:605--659, 1995.

\bibitem{andresen96}
B.~Andresen.
\newblock Finite-time thermodynamics.
\newblock {\em Rev Gen Therm}, 35:647--650, 1996.

\bibitem{thirring70}
W.~Thirring.
\newblock Systems with negative specific heat.
\newblock {\em Z. f. Phys.}, 235:339--352, 1970.

\bibitem{chbihi95}
A.~Chbihi, O.~Schapiro, S.~Salou, and L.G. Sobotka.
\newblock Experimental evidence for a phase transition in nuclear evaporation
  process.
\newblock {\em preprint}, 1995.

\bibitem{lyndenbell95a}
R.M. Lynden-Bell.
\newblock Negative specific heat in clusters of atoms.
\newblock {\em to be published in Galactic Dynamics}, ?, 1995.

\bibitem{lyndenbell99}
D.~Lynden-Bell.
\newblock Negative specific heat in astronomy, physics and chemistry.
\newblock {\em Physica A}, 263:293,http://xxx.lanl.gov/avs/cond--mat/9812172,
  1999.

\bibitem{gross171}
A.~Chbihi, O.~Schapiro, S.~Salou, and D.H.E. Gross.
\newblock Experimental and theoretical search for a phase transition in nuclear
  fragmentation.
\newblock {\em Eur.Phys.J. A}, 5:251--255,
  (1999);http://xxx.lanl.gov/abs/nucl-th/9901016.

\bibitem{schmidt00}
M.~Schmidt, R.~Kusche, T.~Hippler, J.~Donges, W.~Kornm\"uller, B.~von
  Issendorff, and H.~Haberland.
\newblock Negative heat capacity for a cluster of 147 sodium stoms.
\newblock {\em submitted to Nature}, 2000.

\bibitem{dAgostino00}
M.~D'Agostino, F.~Gulminelli, Ph. Chomaz, M.~Bruno, F.~Cannata, R.~Bougault,
  F.~Gramegna, I.~Iori, N.~le~Neindre, G.V. Margagliotti, A.~Moroni, and
  G.~Vannini.
\newblock Negative heat capacity in the critical region of nuclear
  fragmentation: an experimental evidence of the liquid-gas phase transition.
\newblock {\em Phys.Lett.B}, 473:219--225, 2000.

\bibitem{schmidt01}
M.~Schmidt, R.~Kusche, T.~Hippler, J.~Donges, W.~Kornm\"uller, B.~von
  Issendorff, and H.~Haberland.
\newblock Negative heat capacity for a cluster of 147 sodium stoms.
\newblock {\em Phys.Rev.Lett.}, 86:1191--1194, 2001.

\bibitem{gross172}
A.S. Botvina, M.~Bruno, M.~D'Agostino, , and D.H.E. Gross.
\newblock Influence of coulomb interaction of projectile- and target-like
  sources on statistical multifragmentation.
\newblock {\em Phys.Rev.C}, 59:3444--3447, 1999).

\bibitem{gross95}
D.H.E. Gross.
\newblock Statistical decay of very hot nuclei, the production of large
  clusters.
\newblock {\em Rep.Progr.Phys.}, 53:605--658, 1990.

\bibitem{gross150}
D.H.E. Gross, A.~Ecker, and X.Z. Zhang.
\newblock Microcanonical thermodynamics of first order phase transitions
  studied in the potts model.
\newblock {\em Ann. Physik}, 5:446--452, 1996,and
  http://xxx.lanl.gov/abs/cond-mat/9607150.

\bibitem{gulminelli99a}
F.~Gulminelli and Ph. Chomaz.
\newblock Critical behavior in the coexistence region of finite systems.
\newblock {\em Phys.Rev.Lett.}, 82:1402--1405, 1999.

\bibitem{casetti99a}
L.~Casetti, M.~Pettini, and E.G.D. Cohen.
\newblock Geometric approach to hamiltonian dynamics and statistical mechanics.
\newblock {\em cond-mat/9912092}, 1999.

\bibitem{ispolatov01}
I.Ispolatov an~E.G.D.~Cohen.
\newblock On first-order phase transition in microcanonical and canonical
  non-extensive systems.
\newblock {\em http://xxx.lanl.gov/abs/cond-mat/0101311}, 2001.

\bibitem{gross187}
E.V. Votyakov, H.I. Hidmi, A.~De Martino, and D.H.E. Gross.
\newblock Microcanonical mean-field thermodynamics of self-gravitating and
  rotating systems.
\newblock {\em Phys.Rev.Lett.}, 89:031101--1--4;
  http://arXiv.org/abs/cond--mat/0202140, 2002).

\bibitem{gross181}
O.~Fliegans and D.H.E. Gross.
\newblock Effect of angular momentum on equilibrium properties of a
  self-gravitating system.
\newblock {\em Phys.Rev.E}, 65:046143;
  http://xxx.lanl.gov/abs/cond--mat/0102062, 2002.

\bibitem{gross190}
E.V. Votyakov, A.~De Martino, and D.H.E. Gross.
\newblock Thermodynamics of rotating self-gravitating systems.
\newblock {\em submitted to EJPB}, pages
  http://arXiv.org/abs/cond--mat/0207153, 2002).

\bibitem{gross191}
E.V. Votyakov, A.~De Martino, and D.H.E. Gross.
\newblock The Antonov problem for rotating system.
\newblock {\em submitted to Nucl.Phys.A}, pages
  http://arXiv.org/abs/cond--mat/0208230, 2002).

\bibitem{gross177}
D.H.E. Gross.
\newblock What can nuclear collisions teach us about the boiling of water or
  the formation of multi-star systems ?
\newblock In A.~Ventura G.C.~Bonsignori, M.~Bruno and D.~Vretenar, editors,
  {\em Nucleus-Nucleus Collisions Proceedings of the Conference: Bologna 2000
  Structure of the Nucleus at the Dawn of the Century}, The Science and Culture
  Series -- Advanced Scientific Culture, pages 53--60, Bologna, Italy;
  http://arXiv.org/abs/cond-mat/?0006203, 2000. World Scientific.

\bibitem{gross176}
D.H.E. Gross.
\newblock Phase transitions in "Small" systems -- a challenge for
  thermodynamics.
\newblock {\em Nucl.Phys.}, A681:366c--373c, 2001;
  http://arXiv.org/abs/cond-mat/?0006087.

\bibitem{gross189}
D.H.E. Gross.
\newblock Thermo-statistics or topology of the microcanonical entropy surface.
\newblock In T.Dauxois, S.Ruffo, E.Arimondo, and M.Wilkens, editors, {\em
  Dynamics and Thermodynamics of Systems with Long Range Interactions}, Lecture
  Notes in Physics, pages 21--45,cond--mat/0206341, Heidelberg, 2002. Springer.

\bibitem{gross182}
D.H.E. Gross.
\newblock Straight way to thermo-statistics, phase transitions, second law of
  thermodynamics, but without thermodynamic limit.
\newblock {\em http://xxx.lanl.gov/abs/cond-mat/0105313}, 2001.

\bibitem{gross138}
A.~Botvina and D.H.E. Gross.
\newblock Sequential or simultaneous multifragmentation of nuclei.
\newblock {\em Phys.Lett.B}, 344:6--10, 1995.

\bibitem{berdichevsky97}
Victor~L. Berdichevski.
\newblock {\em Thermodynamics of chaos and order}.
\newblock Longman, Edinburgh Gate, Harlow, England, 1997.

\bibitem{yang52}
C.N. Yang and T.D. Lee.
\newblock Statistical theory of equations of state and phase transitions. i.
  theory of condensation.
\newblock {\em Phys. Rev.}, 87:404, 1952.

\bibitem{gallavotti99}
G.~Gallavotti.
\newblock {\em Statistical Mechanics}.
\newblock Texts and Monographs in Physics. Springer, Berlin, 1999.

\bibitem{gross183}
D.H.E. Gross.
\newblock Ensemble probabilistic equilibrium and non-equilibrium thermodynamics
  without the thermodynamic limit.
\newblock In Andrei Khrennikov, editor, {\em Foundations of Probability and
  Physics}, number XIII in PQ-QP: Quantum Probability, White Noise Analysis,
  pages 131--146, Boston, October 2001. ACM, World Scientific.

\bibitem{falconer90}
Kenneth Falconer.
\newblock {\em Fractal Geometry - Mathematical Foundations and Applications}.
\newblock John Wiley \& Sons, Chichester, New York, Brisbane,
  Toronto,Singapore, 1990.

\bibitem{gross186}
D.H.E. Gross.
\newblock Geometric foundation of thermo-statistics, phase transitions, second
  law of thermodynamics, but without thermodynamic limit.
\newblock {\em PCCP}, 4:863--872,http://arXiv.org/abs/cond--mat/0201235,
  (2002).

\bibitem{uffink01}
Jos Uffink.
\newblock Bluff your way in the second law of thermodynamics.
\newblock {\em cond-mat/0005327}.

\bibitem{gross192}
D.H.E. Gross.
\newblock Second law in classical non-extensive systems.
\newblock In D.~Sheehan, editor, {\em Proceedings of the First International
  Conference on Quantum Limits to the Second Law}, University of San Diego,
  2002.

\end{thebibliography}

\end{document}